# Entropic descriptor based reconstruction of three-dimensional porous microstructures using a single cross-section

D. Frączek[a,1], W. Olchawa[b], R. Piasecki[c,*], R. Wiśniowski[d,1],

[a] *Department of Materials Physics, Opole University of Technology, Katowicka 48, 45-061 Opole, Poland*
[b,c,d] *Institute of Physics, University of Opole, Oleska 48, 45-052 Opole, Poland*

## HIGHLIGHTS

- We use a single entropic descriptor to the statistical reconstruction of a 3D porous media.
- All we need to reconstruct an entire structure is its representative cross-section.
- The synthetic starting pattern greatly speeds up the reconstruction process.

## ABSTRACT

The simple entropic method to statistical reconstructing of heterogeneous three-dimensional media from a single two-dimensional image is briefly reported. We apply the entropic descriptor quantifying spatial inhomogeneity that depends on length scale. The multiscale reconstructing procedure is boosted when we start from the synthetic 3D configuration. It is randomly generated with the overlapping spheres of a radius depending on the structure considered. In our scenario, the simulated annealing terminates when all assigned temperature loops are completed. Thus, the quality of the results for different samples can be fairly compared. The reliable 3D reconstructions of porous sandstone, ceramics and carbonate samples are obtained. They suggest the entropic method is a promising approach, which offers a kind of compromise between the computational efficiency and the accuracy of the statistical reconstructions. Furthermore, this approach is versatile enough to be extended for any multiphase medium.

(Some figures in this article are in colour only in the electronic version)

## 1. Introduction

It is commonly accepted that three-dimensional microstructural information is significant for modelling effective properties of random heterogeneous materials since the physical properties evidently are correlated with its microstructure [1, 2]. Recently, it has been pointed the disordered materials can be divided into two wide classes: spatially stationary and spatially nonstationary media [3]. The former class, for which "the probability distribution function of any property does not change when shifted in space" [3], belong to the field of our interest. In such a class, a representative sub-domain usually serves as a target pattern for the preferred here group of stochastic reconstruction methods, which allow incorporating an arbitrary number of correlation functions or specific descriptors of any types. Thus, the developing of spatial descriptors that provide quantitative information about some features of configuration of phases is of some importance.

---

* Corresponding author. Tel.: +48 77 452 7285; fax: +48 77 452 7290.
  E-mail addresses: d.fraczek@po.opole.pl (D. Frączek), wolch@uni.opole.pl (W. Olchawa), piaser@uni.opole.pl (R. Piasecki), ryszard.wisniowski@gmail.com (R. Wiśniowski).

[1] D. Frączek and R. Wiśniowski are recipients of a Ph.D. scholarship under a project funded by the European Social Fund.

On the other hand, the important part of disordered media are the so-called multiscale complex systems. They show characteristic structural features at different length scales. It makes the statistical reconstruction within the popular simulated annealing (SA) approach, a highly non-trivial task even for a 2D image, e.g., laser-speckle pattern by mean of the two-point correlation function $S_2$ alone [4]. Even the pair of $S_2$ and cluster correlation function $C_2$, cannot capture all characteristic morphological features of given concrete sample cross-section with the irregularly shaped stone phase [5]. On the other hand, a distinct hybrid approach was developed for the statistical reconstruction [6]. It employs the so-called entropic descriptors (EDs) averaged over sampling cells, $S_\Lambda$ for *spatial inhomogeneity* [7], $C_{\lambda, S}$ for *spatial statistical complexity* [8] and their grey-level counterparts, $G_\Lambda$ and $C_{\lambda, G}$ [9]. This entropic method can be also applied to another complex class like labyrinth patterns [10]. It is worth noticing the limited structural information provided by a pair of correlations functions, e.g. $S2$-$C_2$ and a pair of EDs, e.g. $S_\Lambda$-$C_{\lambda, S}$}, is comparatively different. This observation allowed the introducing of a weighted doubly-hybrid (WDH) approach. This is a convincing method of the statistical reconstruction of any two-phase patterns of islands of miscellaneous shapes and poly-dispersed in sizes [11]. It should be noticed that a recently devised dilation–erosion method generalizing the $S2$-$C_2$ approach [12], can be successfully extended to reconstructing of topologically complex three-dimensional microstructure from 2D micrograph [13].

One remark more is in order here. The aforementioned concrete cross-section comprising stone aggregates, which are relatively big compared to the pattern size and of size larger than the average spacing between clusters, is not fully the representative sample. Even though, the reconstructions obtained by the different approaches, cf. Fig. 2d in [5], inset of Fig. 6b in [11] and Fig. 4 in [14], are qualitatively similar in shapes of aggregates. However, they differ in the number of clusters, their areas and interface values (unpublished analysis). In the latter work, a cross-correlation function containing global multiple-point information is applied [15]. The further extension of that approach capable of integrating hard and soft data into the simulation and reconstruction process is also discussed [16]. This kind of multiple-point statistics methodology was originated a long ago [17] and plenty modifications of training image-based stochastic models with various sampling procedures have been developed, see e.g., [18, 19].

However, this work focuses on the entropic approach that belongs to still vital stochastic reconstruction methods. Here, we present the first three-dimensional application of an earlier developed entropic method (for 2D patterns) to reconstructing 3D porous microstructures from a limited morphological information contained in a single cross-section. The obtained results suggest that its simplest version is a good compromise between the computational efficiency and the quality of the statistical reconstructions. However, some of spatial features detected by the applied entropic descriptor $S_\Lambda$, can be treated as complementary ones to those revealed within the abovementioned different developments.

## 2. The entropic cost function

The basis of our entropy method of statistical reconstruction is the assumption that scale sensitive statistical properties of a microstructure can be described, at least in some part, by means of specific entropic descriptors (EDs). In this project, we would like to employ the simplest approach that uses only a one entropic descriptor that is the spatial inhomogeneity measure $S_\Lambda(k)$ as a function of length scale $k$ for a given 2D binary pattern of size $L{\times}L$ (in pixels),

$$S_\Lambda(k) = \frac{1}{\lambda} \left[ S_{max}(k) - S(k) \right].$$ (2.1)

This ED makes the use of micro-canonical current entropy, $S(k) = k_B \ln \Omega(k)$, and its maximum, $S_{max}(k) = k_B \ln \Omega_{max}(k)$, where conveniently Boltzmann's constant can be put equal to unity. The $\Omega(k)$ and $\Omega_{max}(k)$ correspond to the number of microstates realizing the current and the most uniform configurational macrostates properly defined at every discrete length scale $k$. This scale is given by the side length of the sampling cell of size $k \times k$ and sliding by a unit distance. Black pixels are treated as finite size, $1 \times 1$, unit objects. For instance, at given scale a current macrostate is described by a set of $i$th cell occupation numbers $\{n_i(k)\}$ by black pixels, $i = 1, 2, \ldots, \lambda(k)$. The number of allowed positions for the sliding cell equals $\lambda(k) = [L - k + 1]^2$. More detailed description of this and other EDs can be found in [6]. As it becomes clear later on, there is no need to consider three-dimensional counterpart of the ED since in the present project the only available morphological information is contained in a single 2D cross-section of a porous 3D sample.

The proposed quantitative measure for spatial inhomogeneity averaged over the number of sampling cells, can be naturally applied to evaluation of statistical similarity of any two structures, say 'A' and 'B'. The more statistically similar structures 'A' and 'B' are, the closer are values of the corresponding curves $S_\Delta(k; A)$ and $S_\Delta(k; B)$, and reversely. The statistical „distance" per cell between such two curves can be calculated as a sum over length scales of the squared differences, $[S_\Delta(k; A) - S_\Delta(k; B)]^2$. It should be underlined here, the information contained in the $S_\Delta(k)$-curve alone, is sufficient to perform the "inverse" reconstruction without of the knowledge of a target pattern. Thus, in the entropic descriptor approach there is no need to use "training" image techniques, typical for the MPS methodology. Our approach making use of EDs has been already applied to the statistical reconstruction of 2D complex labyrinth patterns [10], islands poly-dispersed in size [11]. Here, we apply the ED based method for the first time to statistical reconstructions of three-dimensional, $L \times L \times L$, porous samples. In addition, the task should be completed on the basis a one 2D input image only.

The general idea of our approach is quite simple. Let us introduce a Cartesian coordination system with the origin in a corner of the cube and axes oriented along its edges. The sample is treated as a set composed of three subsets each of $L$ planes. The three subsets contain the stacks of the $L$ planes being cross-sections of the 3D sample perpendicularly to the $x$, $y$ and $z$-axis, respectively. Our final 3D reconstruction is acceptable when any plane of this set is statistically similar to the 2D input image treated as the target pattern. To be precise, we define the entropic cost function per plane, i.e. the averaged objective function

$$E_{avg} = \frac{1}{3L} \sum_{p=1}^{3L} E_p \, . \tag{2.2}$$

Here, $E_p$ denotes the sum of squared and normalized differences between the values of normalized EDs related to a current configuration of plane $p$ and the target pattern. The latter can be also selected from a larger parent image as a representative sub-domain (this is a case here). Then, the sum is averaged over the number $N_k$ of the length scales considered,

$$E_p = \frac{1}{N_k} \sum_{k=k_0}^{k_1} \left[ \widetilde{S}_\Delta(k, p) - \widetilde{S}_\Delta^T(k) \right]^2 \, . \tag{2.3}$$

The EDs are normalized with regard to the maximal value of target entropic descriptor, $S_\Delta^T(k_{max})$, marked with the superscript 'T'

$$\widetilde{S}_\Delta(k, p) - \widetilde{S}_\Delta^T(k) \equiv [S_\Delta(k, p) - S_\Delta^T(k)] / S_\Delta^T(k_{max}) \, . \tag{2.4}$$

The maximal value is reached at a scale $k_{max}$.

The further stages of our approach are presented below. Generally, during the reconstruction process for the chosen number of loops with the assigned increasing lengths, the value of the cost function $E_{avg}$ considerably decreases. The simulated annealing scenario for the temperature loops ensures the proper limiting behaviour of minimized $E_{avg}$. At the same time, also the interface $I$ per plane denoted here as the $<I>$ is actively monitored.

## 3. The multi-scale entropic reconstruction

### 3.1 The selection process of a representative subdomain

Each of the representative subdomain of size $L \times L$ with $L = 300$ is chosen from the corresponding larger 2D parent image (sample cross-section). These 2D parent images of isotropic porous samples (sandstone of size $700 \times 700$, ceramics and carbonate of sizes $500 \times 500$) were provided within the voluntary project originated by the people of CSIRO*. Firstly, using the $L \times L$-sampling cell, we detect for each of the parent images on what length scale $k \equiv k_{max}$, the most frequently appears the first peak of $S_A(k)$ that usually is also a global maximum. Then, the list of the locations of the corresponding sampling cells is sorted over the slightly fluctuating volume fractions of the phase of interest. Among the cases with the volume fraction closest to the parent image, we choose a one, for which also the value of a plane "interface" is proportional to that of the parent image. As a result, we obtain a representative 2D target pattern of size $300 \times 300$ with the proper length scale $k_{max}$, the appropriate phase volume fraction and the corresponding interface value. Now, the 2D target $S_A^T(k)$ - curve, as a function of length scale $k$ can be calculated.

### 3.2 Preparation of a preliminary 3D configuration

It may be convenient to start with the reversed phase colours in the 2D target image. So, for the present samples, the volume fraction of black phase after the reversing of colours, is always less than 0.5. Now, let us consider 3D cube of size $L^3$ composed of only black phase voxels $1 \times 1 \times 1$. To generate an initial random 3D configuration with the needed volume fraction of black phase, instead of white single voxels we use the overlapping spheres composed of white voxels and having a fixed radius $R$. The positions of sphere centres are drawn with a uniform probability distribution inside the cube and in the external zone of an appropriate width. The width of the zone is determined in such a way that at least one voxel of every white sphere must be an internal voxel of the cube. Close to the ending of cutting white wholes (or pieces) from the black 3D matrix, some trials may be rejected until the same volume fraction of black phase as in the target is obtained. This way of porous configuration generating can be named as the balls-procedure (BP).

The entropic cost function, $E_{avg}$, described by (2.1-4) shows a feature that is particularly useful for the statistical reconstruction purposes. Let us generate by BP trial initial 3D configurations for a series of discrete values of $R$ taken from a wide enough interval. Then, for the associated family of $E_{avg}$-curves the approximate local minimum of $E_{avg}$ appears for a characteristic discrete value of $R$. In this way, the optimal starting 3D configuration can be prepared immediately in a few seconds using the detected $R$-value.

In fact, at this stage a coarse 3D reconstruction of interest is received. For instance, the corresponding initial value $E_{avg}$(start) is less than $68.8 \times 10^{-3}$ for sandstone, $49.0 \times 10^{-3}$ for ceramics and $86.1 \times 10^{-3}$ for carbonate. Since our algorithm is the most efficient in creating aggregates, a higher value of the initial interface compared to the target one is preferred here. The further work is done making use of Monte Carlo (MC) simulated annealing. We point out that only limited number of scales $k$ can be taken into account. Within the present approach, we use every second scale, $k = 2, 4, \ldots$, until the half of the image size. Two reasons are for that. First, we are interested in morphological features, which are typical for rather small length scales. Similar range of scales is characteristic for other methods, e.g., for two-point correlation functions [1]. On the other hand, the computations performed for 75 scales instead of 150 is obviously being much more computationally efficient and still satisfactory enough as well.

### 3.3 Simulated annealing

At this stage, we employ the MC simulated annealing, which should further minimize the entropic cost function, $E_{avg}$(start). After the interchange of the voxels (here equivalently one can say: pixels on planes), the new trial configuration equivalently called the system's state, is accepted with probability $p(\Delta E_{\text{avg}})$, according to the Metropolis-MC acceptance rule

$$p(\Delta E_{avg}) = \min[1, \exp(-\Delta E_{avg}/T)]. \tag{3.1}$$

Here, $\Delta E_{avg} = E_{avg,\,new} - E_{avg,\,old}$ is the difference in "energy" between two successive states that is related to the changes on 6 planes each time. Upon acceptance, the trial pattern becomes a current one, and the evolving procedure is repeated. A fictitious temperature $T$ follows the cooling schedule, $T(l)/T(0) = \gamma$, with the chosen parameter $\gamma = 0.80$, the initial temperature $T(0) = 10^{-8}$, the $l$th temperature loop of increasing length and fixed number of the loops, $l = 26$.

However, some reconstruction details are the non-standard. Having determined the "worst" $\omega$-plane with the maximal energy $E_p$ among $3L$ planes, we are in position to start the preferential selection of two voxels of different phases, called here as "biased mode". If the volume fraction of black phase on $\omega$-plane is higher (lower) than of the target one, then a voxel drawn should be of black (white) colour. To accelerate computation, the voxel of the remaining colour is drawn in such a way that it does not belong to the three planes connected with the first voxel. In addition, for symmetry reasons, at least for a one of the three planes associated with the second voxel, the volume fraction should change toward a target value, too.

Let us denote the numbers of black n.n. and black n.n.n. for a white centre as $w_{nn}$ and $w_{nnn}$, and similarly, for a black centre as $b_{nn}$ and $b_{nnn}$. By treating n.n. on an equal footing with n.n.n., one can ensure their equal contributions. Thus, the appropriate weights are introduced in the "neighbouring" rules for every two pixels of different phases randomly selected:

$$(10b_{nn} + 3b_{nnn} < 10w_{nn} + 3w_{nnn}) \quad \text{and} \quad (b_{nn} \leq w_{nn}) \tag{3.2}$$

or

$$(10b_{nn} + 3b_{nnn} = 10w_{nn} + 3w_{nnn}) \quad \text{and} \quad (b_{nn} < w_{nn}). \tag{3.3}$$

At this stage, our algorithm by creating aggregates favours the lowering of the averaged interface, $<I>$. When the current value of the $<I>$ is below of the $I_{\text{target}}$, then the rules (3.2-3) are not active. Then, the entirely random selection of two voxels of different phases, called here as "unbiased mode", favours the raising of the $<I>$ value. Thus, we apply the following switching: when the current $<I>$ value exceeds the value of $I_{\text{target}}$, the biased mode come into play while in the opposite case, the unbiased one. If all temperature loops are completed then the reconstruction terminates. The method is tested in next section on three 2D single cross-sections for 3D different porous microstructures.

## 4. Examples of Monte Carlo simulations

For each of the 2D parent images of isotropic porous sandstone, ceramics and carbonate samples, the earlier selected as the target patterns the representative subdomains of size $L \times L$ with $L = 300$ and porous phase fraction $\phi = 0.19731, 0.38144$ and $0.14381$, respectively, were the only allowable input to reconstruct the needed 3D structures. Here, the corresponding target curves suggest that the sandstone is the most representative sample because of its spatial uniformity. On the other hand, the carbonate is the worst as confirms also a simple observation by a naked eye. Surprisingly, the obtained results indicate the ceramics as the most difficult sample to reconstructing by our method.

However, even using the simplest version of the entropic approach within the same simulated annealing scenario, the obtained results are quite satisfactory. As it can be seen in Table 1, the corresponding outcomes differ in the final ratio $E_{avg}(\text{start}) / E_{avg}(\text{end})$ as well as in the numbers of accepted MC-steps.

Table 1. Some of the results for the entropic $S_A(k)$-descriptor based the multiscale statistical reconstructions of 3D porous samples from the related single cross-sections.

| Sample | $E_{avg}(\text{start})$ | $E_{avg}(\text{end})$ | $E_{avg}(\text{start}) /$ $E_{avg}(\text{end})$ | # of accepted MC steps |
|---|---|---|---|---|
| Sandstone | $68.8 \times 10^{-3}$ | $0.161 \times 10^{-3}$ | 429 | $1100 \times 10^3$ |
| Ceramics | $49.0 \times 10^{-3}$ | $1.330 \times 10^{-3}$ | 37 | $1400 \times 10^3$ |
| Carbonate | $86.1 \times 10^{-3}$ | $0.557 \times 10^{-3}$ | 155 | $1300 \times 10^3$ |

For the CPU Intel 7 (3.3 GHz) without code parallelization, the overall computation time for sandstone, ceramics and carbonate samples was about 4.3 h, 5.4 h and 5 h, respectively. In Figs. 1a, 2a and 3a the EDs solid curves (red online) correspond to the proper target pattern, see the upper inset, while open circles (blue online) refer to the plane, see the bottom inset, being a one of the 900 planes with the $E_p$ energy, which is the nearest to the final $E_{avg}$ energy (after the 3D reconstruction). In turn, for illustration purposes, in Figs. 1b, 2b and 3b the 3D exterior views of the reconstructed samples (R#sand, R#ceram and R#carb) are presented.

Bearing in mind that our method was primarily developed to materials composed of solid phases, here it has been applied to the porous media. Nevertheless, among the final reconstructions, the fraction of isolated solid clusters in the carbonate sample was of order $10^{-4}$ while for the remaining two cubes the related fractions are two orders lower, yet. However, the program current version can be easily modified to avoid those unrealistic effects.

On the other hand, we have checked the possible impact of isolated solid clusters making use of the simple algorithm. The main point is how to consolidate the black phase preserving the overall isotropy of the samples. This condition can be fulfilled by selecting randomly a one of sixth main directions in order to make a shift of a given isolated cluster. As expected, the values of entropic descriptor $S_A(k; 3D)$ calculated for each of the final 3D configurations without of any isolated cluster are practically identical with the counterparts referred to the reconstructed cubes (R#sand, R#ceram and R#carb).

In addition, the method of multi-scale entropic reconstruction can be enriched by considering also other entropic descriptors. This could allow further improving the accuracy of the 3D reconstructions. It is worth noticing that for a more general three-phase microstructure reconstruction in 3D we need, at least two entropic descriptors obtained by the phase separated entropic measure [20].

## 5.  Conclusions

We proposed and developed a method based on the single entropic descriptor with the purpose of reconstructing of entire 3D porous medium from a one cross-section. Given a sample cross-section is treated as 2D input image (target pattern) and provides the corresponding target curve for the chosen entropic descriptor. The proposed average entropic cost function is optimized within the simulated annealing scenario. The beginning from a synthetic initial 3D configuration allows boosting the multiscale reconstructing process. The 2D input images of porous sandstone, ceramics and carbonate are used to verify the entropic method. The final 3D reconstructions have a property: the entropic descriptor curve associated with any cross-section of the cube runs very close to the corresponding target curve. The obtained results suggest the entropic method offers a kind of compromise between the computational efficiency and the acceptable accuracy of the statistical reconstructions. Our approach can be also extended to any multiphase medium.

**Figure captions**

**Fig. 1.** The quality of the entropic $S_\Delta(k)$-descriptor based method of reconstructing of three-dimensional porous (white phase) microstructures from a single sample cross-section. a) Comparison of the entropic descriptor functions for the target image (solid line) of sandstone sample and for a one of 900 planes in the reconstructed cube (R#sand, open circles). This selected plane is the optimal one since the associated $E_p$ energy is the nearest to the final $E_{avg}$. In the upper inset the 2D target image is depicted, while the bottom inset refers to the selected optimal plane; b) 3D exterior view of the reconstructed cube (R#sand).

**Fig. 2.** Same as in Fig. 1 but for ceramics sample.

**Fig. 3.** Same as in Fig. 1 but for carbonate sample.

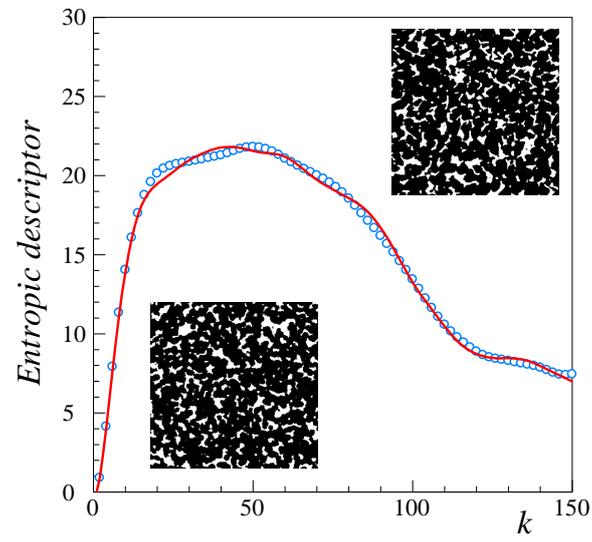

**Fig. 1a**

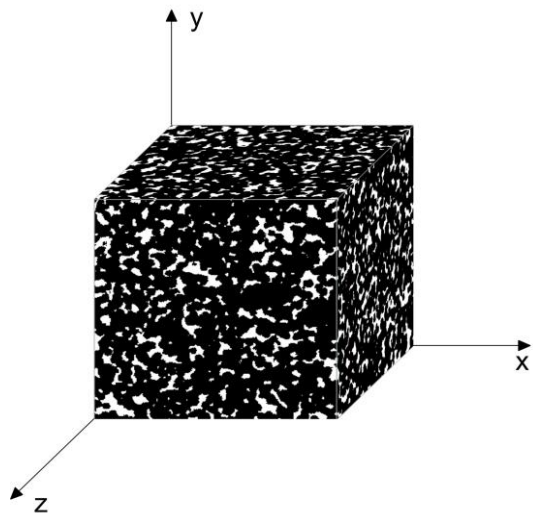

**Fig. 1b**

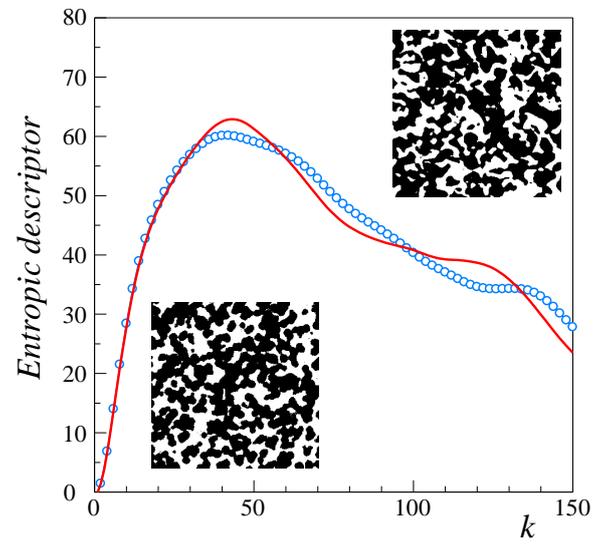

**Fig. 2a**

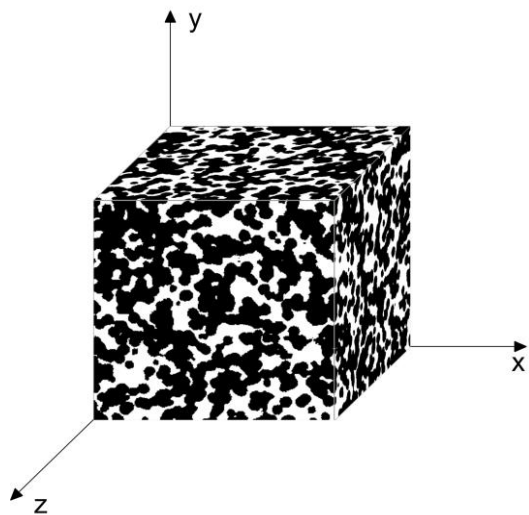

**Fig. 2b**

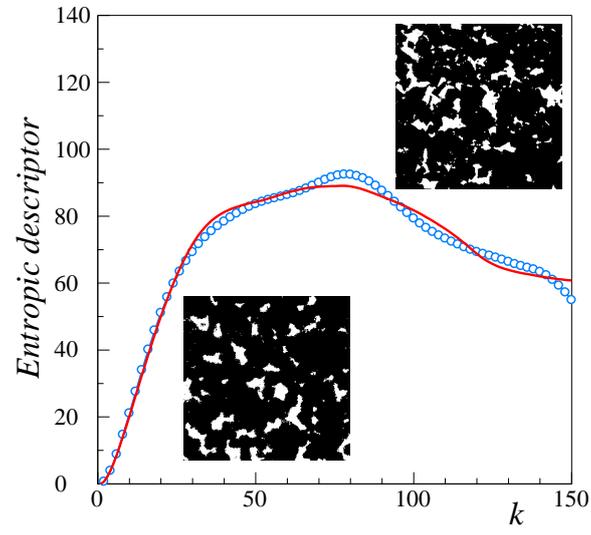

**Fig. 3a**

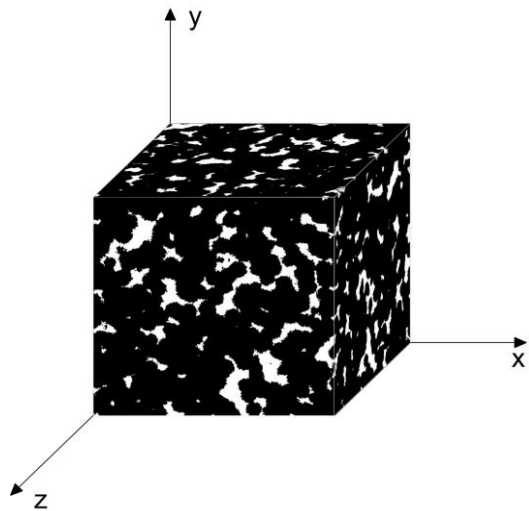

**Fig. 3b**